\documentclass[tighten,twocolumn]{aastex62}
\usepackage{graphicx,natbib,amsmath,xfrac}
\bibpunct{(}{)}{;}{a}{}{,}
\let\tmpclearpage\clearpage
\let\clearpage\relax
\DeclareMathOperator\erf{erf}

\newcommand{\narrowfig}[3]{%
\begin{figure}[tb!]
\begin{center}
\includegraphics[width=3.39375in]{#1}
\caption{#3}
\label{#2}
\end{center}
\end{figure}
}

\newcommand{\widefig}[3]{%
\begin{figure*}[tb!]
\begin{center}
\includegraphics[width=7.1in]{#1}
\caption{#3}
\label{#2}
\end{center}
\end{figure*}
}

\journalinfo{\doi{10.3847/1538-3881/aaedc0}}

\begin{document}

\title{Characterization of Cameras for the COSMO K-Coronagraph}
\shorttitle{Characterization of Cameras for the COSMO K-Coronagraph}
\author[0000-0002-5084-4661]{A.G.~de~Wijn}
\altaffiliation{The National Center for Atmospheric Research is sponsored by the National Science Foundation.}
\affiliation{High Altitude Observatory, National Center for Atmospheric Research, P.O. Box 3000, Boulder, CO 80307, USA}
\email{dwijn@ucar.edu}
\shortauthors{De Wijn}

\begin{abstract}

	Digital image sensors are ubiquitous in astronomical instrumentation and it is well known that they suffer from issues that must be corrected for data to be scientifically useful.
	I present discussion on errors resulting from digitization and characterization of nonlinearity and ADC errors of the PhotonFocus MV-D1024E cameras selected for the K-Coronagraph of the Coronal Solar Magnetism Observatory.
	I derive an analytic expression for quantization errors.
	The MV-D1024E camera has adequate bit depth for which quantization error is not an issue.
	I show that this is not the case for all cameras, particularly those with deep wells and low read noise.
	The impact of nonlinearity and ADC errors on science observations of the K-Coronagraph is analyzed using a simplified telescope model.
	Errors caused by the camera ADCs result in systematic errors in the measurement of the polarimetric signal of several times $10^{-9}~B_\odot$, which is
	about an order of magnitude above the desired sensitivity.
	I demonstrate a method for post-facto data correction using a lookup table and derive parameters from camera characterization measurements that were made with a lab setup.
	Nonlinearity is traditionally addressed with a global correction.
	I show through analysis of calibration data that for the MV-D1024E this correction leaves residual systematic errors after dark and gain correction of up to $1\%$ of the signal.
	I demonstrate that a pixel-wise correction of nonlinearity reduces the errors to below $0.1\%$.
	These corrections are necessary for the K-Coronagraph data products to meet the science requirements.
	They have been implemented in the instrument data acquisition system and data reduction pipeline.
	While no other instruments besides the K-Coronagraph or cameras besides the MV-D1024E are discussed here, the results are illustrative for all instruments and cameras.
\end{abstract}


\section{Introduction}\label{sec:introduction}

Observers have always sought to record their observations in the most accurate way available to them.
Early observers made drawings by hand.
The first photograph of the sun was made by Fizeau and Foucault in 1845 using the Daguerreotype process.
Nowadays, most imaging of the sun is done using digital image sensors.
The dominant types of digital image sensors are the charge coupled device (CCD) and the active pixel sensor (APS).
The latter is based on the Complementary Metal Oxide Semiconductor (CMOS) process and also frequently referred to as a CMOS sensor.

The CCD was invented by \cite{1970BSTJ...49..587B}.
It consists of a grid of pixels that can accumulate an electrical charge.
Electron-hole pairs are generated by photons that are absorbed in a positively doped semiconductor layer, and the electrons are trapped in a potential well created by a positively biased electrode.
The number of electrons that can be stored this way in a pixel is referred to as the full-well capacity.
The device can shift the electrons from one well to another by selectively biasing neighboring electrodes, so that the electrical charge can be transported to the edge of the light-sensitive area.
There, external electronics convert the charge to a voltage.
This voltage, in turn, can be read out and digitized by an analog-to-digital converter (ADC).

The APS was developed around the same time \citep{1968ITED...15..202N,1969IJSSC...4..333C,1969ITED...16..240W}.
The APS differs from the CCD sensor in that the charge-to-voltage conversion and amplification is performed by electronics integrated in each pixel.
Uniformity of the signal over the sensor is lower than it is for a CCD because each pixel charge is converted by different electronics.
Early APS sensors were inferior to CCD sensors due to limitations in CMOS fabrication processes, and the technology was largely abandoned until improvements in photolithography made it possible to produce usable sensors.

APS sensors have several disadvantages in addition to the issue of non-uniformity.
Because some electronics are integrated with each pixel, the photosensitive area is reduced in size compared to a CCD, resulting in a commensurate loss of efficiency and reduced well capacity.
Many sensors now use lenslet arrays that mitigate the issue of efficiency to some extent.
Back-side illuminated sensors do not suffer from this issue and have other desirable properties.
They are becoming more common despite being more difficult to manufacture.
The production of APS sensors is also tuned to devices that operate in the visible range of the spectrum.
Sensors must have a thicker photon absorption region in order to have good sensitivity in the near-infrared, which requires other modifications such as a higher pixel bias voltage, which in turn affect the integrated analog and digital circuits.
Consequently, APS sensors are generally much less sensitive than CCD sensors specifically designed for operation in the near-infrared.

However, APS sensors also offer several advantages over CCD sensors.
Their integrated electronics process the charge on pixels in a massively parallel fashion, resulting in the possibility of high-speed readout of images.
The ability to integrate complicated circuits in addition to the charge-to-voltage circuit included in each pixel on the same substrate also results in lower power consumption, and lower fabrication and integration costs.
The quality of APS sensors has continued to improve, largely due to high-volume applications such as cell phone cameras.
APS sensors have increased in popularity for scientific observations although CCD sensors continue to exhibit properties superior to APS sensors in terms of uniformity, sensitivity, and noise.
For solar observations they are used in applications that use polarimetry because high frame rates are desired to reduce seeing-induced cross-talk \citep{2012ApJ...757...45C}.

The K-Coronagraph instrument that resides at NCAR's Mauna Loa Solar Observatory and is an element of the Coronal Solar Magnetism Observatory (COSMO).
It uses polarimetry in a white-light band around $735~\mathrm{nm}$ to discern the linearly polarized K-corona from a much brighter sky background.
Linear polarization is encoded into intensity signals using an electro-optical modulator and polarizer, and later recovered by adding appropriately scaled linearly independent measurements \citep{2013ApJ...774...85H}.
The requirement on the imaging system is to reach a sensitivity of $10^{-9}~B_\odot$ in $15~\mathrm{s}$ to permit the detection of faint structures in the corona and capture the dynamics of fast coronal mass ejections.
A high signal-to-noise ratio is additionally needed to separate the K-corona from the much brighter background sky.
Hence, the driving factor in the search for a camera was the product of full-well capacity and frame rate.
The PhotonFocus MV-D1024E camera utilizing the PhotonFocus A1024B CMOS APS sensor with high full-well capacity of about $180\times10^3~\mathrm{e}^-$ and capable of high frame rate of up to $150~\mathrm{Hz}$ was selected.
Many cycles of the modulator must be averaged in order to reach the desired sensitivity because it is much less than the least significant bit of the camera readout or data number (DN).
However, a bias is introduced in the average by the discretization performed by the camera, which is called \emph{quantization error}.

The A1024B sensor is intended for industrial vision, e.g., for high-speed product inspection.
It was initially evaluated for suitability in a scientific instrument by \cite{cosmotech17}.
They concluded that nonlinearity and high read noise were a concern, but found no reason the camera could not be used.
During those tests it was also noticed in a histogram of readout values that some values would occur more frequently than others.
I will refer to this as \emph{DN error} because it is caused by differences in the equivalent values of a DN.
In essence, each DN is not the same size as the preceding one.

Nonlinearity, i.e., that the readout value is not directly proportional to the number of incident photons, is a well-known issue in many cameras.
Cameras have many sources of nonlinearity, such as the response of the photosensitive area and the electronic circuits employed in current-to-voltage conversion and amplification.

Quantization and DN errors are a concern for polarimetry because it requires taking the difference between two large intensity values to find a small signal, often with precision much better than the least significant bit in the readout.

In this manuscript I will derive an analytic expression for the quantization error.
Error resulting from DN error and nonlinearity will be discussed in the context of characterization and correction.
After a brief description of the calibration setup I will present a model of the instrument that shows the effect of DN error on measurements of the K-Coronagraph, and discuss a method for post-facto mitigation.
The last part of the paper describes the correction of sensor nonlinearity.

\section{Quantization}\label{sec:discretization}

\narrowfig{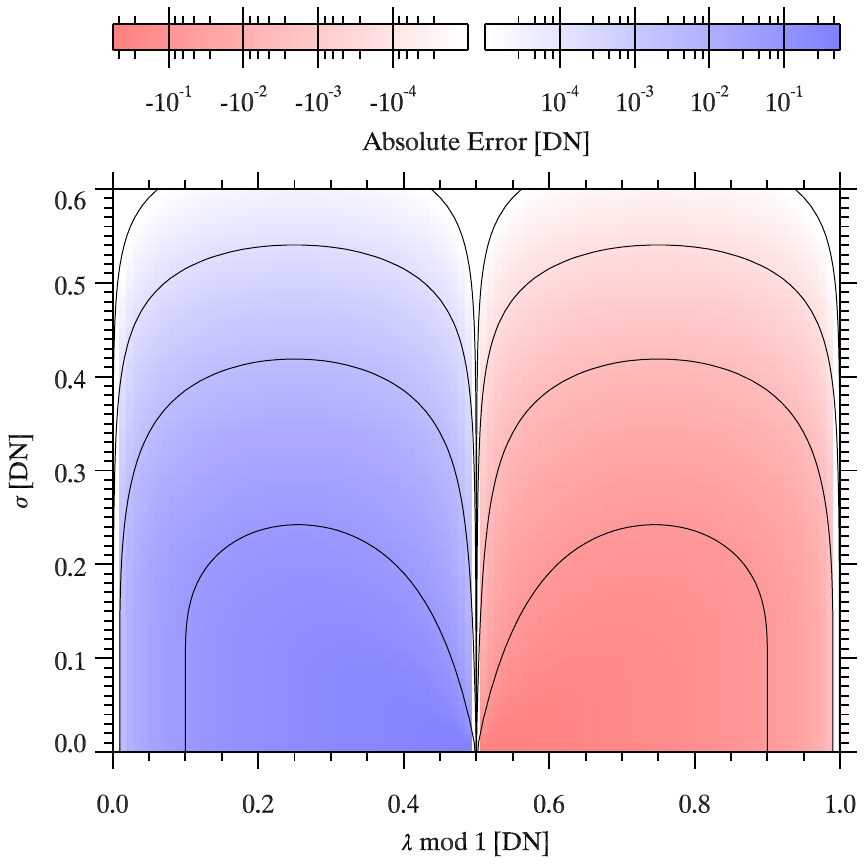}{fig:discretization}{Systematic error in the averaged readout value due to quantization error as a function of random error.
Contours in the figure correspond to $10^{-1}$, $10^{-2}$, $10^{-3}$, and $10^{-4}~\mathrm{DN}$ systematic error.
Low amounts of random noise result in large systematic errors.}

The voltage associated with the pixel is digitized by an ADC, which implies a discretization step.
In principle the discretization could resolve a single photoelectron, but in case of the PhotonFocus A1024B sensor this would require an 18-bit ADC, and the camera only has a 12-bit ADC.
The light transfer test by \cite{cosmotech17} found a gain factor of $44~\mathrm{e}^{-}/\mathrm{DN}$.
Consequently if the noise on a measurement is substantially below $44~\mathrm{e}^{-}$ the readout value will be deterministic and always return the same value, resulting in up to $0.5~\mathrm{DN}$ systematic error on the average.
Therefore, an adequate amount of random noise must be present in order to average below $1~\mathrm{DN}$.

Adding random noise removes the deterministic behavior of the ADC so that it becomes possible to average many measurements and find the true value with sub-DN accuracy.
Counterintuitively this means that adding random noise to a measurement improves the accuracy, but this improvement comes at the cost of an increase in the number of samples required to reach a given precision.
The quantization error is given by
\begin{equation}\label{eq:delta}
	\delta = \lambda - \sum_{k=1}^{N-1} k \, p_k\\
\end{equation}
where $\lambda$ is the true signal mean, $p_k$ is the probability of measuring the value $k$, and $N-1$ is the maximum readout value.
We can write $p_k$ in terms of the cumulative distribution function of the signal $F(x)$,
\begin{equation}
	p_k =
	\begin{cases}
		F(\tfrac12) & : k=0\\
		F(k+\tfrac12) - F(k-\tfrac12) & : 0<k<N-1\\
		1 - F(N-\tfrac32) & : k=N-1
	\end{cases}
\end{equation}

In most practical situations the noise distribution can be approximated by a normal distribution,
\begin{equation}\label{eq:normal}
	F(x)=\frac12\left[1+\erf\left(\frac{x-\lambda}{\sqrt{2}\sigma}\right)\right]
\end{equation}
and $\lambda$ will not be close to either $0$ or $N-1$.
Readout values more than a few standard deviations $\sigma$ away from $\lambda$ are exceedingly unlikely to occur and do not meaningfully contribute to the average.
Let us now choose $k_1$ and $k_2$ that satisfy that condition, i.e., $F(x)\approx0$ for $x<k_1$ and $F(x)\approx1$ for $x>k_2$, and also $k_1+k_2=2\lfloor\lambda\rfloor$.
We can then approximate eq.~\ref{eq:delta} by limiting the summation between $k_1$ and $k_2$ to find
\begin{eqnarray}
	\delta&\approx&\lambda - k_2 + \sum_{k=k_1}^{k_2-1} F(k+\tfrac12)\nonumber\\
	&{}={}&\lambda' - \frac12\left[1 - \sum_{k=k_1-\lfloor\lambda\rfloor}^{k_2-\lfloor\lambda\rfloor}\erf\left(\frac{k+\tfrac12-\lambda'}{\sqrt{2}\sigma}\right)\right]
\end{eqnarray}
with $\lambda'=\lambda\mod1$.
The magnitude of the systematic error as a function of $\lambda'$ and $\sigma$ is shown in Fig.~\ref{fig:discretization}.

The science requirements of the K-Coronagraph drive a need for high photoelectron flux, which can be realized through high frame rate, deep wells, or a combination of the two.
However, it is now clear that the camera selected must have adequate noise to satisfy the accuracy requirement.
The science requirements state the need for a dynamic range of $10^4$.
With a 12-bit camera and using a factor of 2 margin it must be possible to average the data to an absolute error of $0.2~\mathrm{DN}$ in order to satisfy the requirement.
The figure shows that $0.2~\mathrm{DN}$ random error is adequate to meet this requirement.
The read noise of the PhotonFocus MV-D1024E is large at about $182~\mathrm{e}^-$ compared to the equivalent value of 1~DN of about $44~\mathrm{e}^-$ \citep{cosmotech17}, so that quantization error is not a concern.

The MV-D1024E camera has deep wells and a relatively slow frame rate.
Cameras that reach a similar rate of photoelectrons accumulated through high frame rates instead can often only reach those frame rates when an 8-bit image is read out due to bandwidth constraints.
For instance, the Mikrotron EoSens CL has a full well capacity of $30\times10^3~\mathrm{e}^-$ and can operate at frame rates in excess of $500~\mathrm{Hz}$ at 8-bit resolution.
However, it must now be possible to average down to at least $0.01~\mathrm{DN}$, which requires about $0.4~\mathrm{DN}$ random error.
The sensor in this camera has a read noise of about $37~\mathrm{e}^-$, which equates to about $0.3~\mathrm{DN}$, meaning that additional noise from other sources (e.g., photon noise) is needed to meet the accuracy requirement.

\section{DN error and nonlinearity Characterization Setup}\label{sec:setup}

A laboratory setup was used for characterization of DN error and nonlinearity.
It is shown schematically in Fig.~\ref{fig:setup}.
A collimated beam of light is produced from a computer-controlled ThorLabs M735L2 LED that emits light in a broad peak around $735~\textrm{nm}$.
This light source was chosen for its reasonably close match to the design wavelength of the K-Coronagraph.
A piece of uncoated glass reflects about $10\%$ of the light toward a ThorLabs SM1PD1A photodiode that will be used as a reference.
Photodiodes are well-known to exhibit linear response over several decades \citep[e.g.,][]{1979ApOpt..18.1555B}.
The photodiode signal is amplified with a Terahertz Technologies Inc.\ PDA-750 amplifier, the output of which is digitized and recorded.
The arm with the photodiode has an aperture stop, an RG695 colored glass filter, and an ND $1.0$ filter.
A lens focuses the light on the photodiode.
The light transmitted through the plate glass beamsplitter passes through a ground-glass diffuser prior to going through an RG695 filter and falling onto the camera sensor.
The RG695 filters close off lens tubes in order to reject ambient light.
The setup is controlled through a LabView program and is automated to capture data sets specifically designed for calibration of the DN error or nonlinearity.
The K-Coronagraph is a dual-beam polarimeter that uses two cameras.
Two sets of cameras have been used so far.
All cameras were calibrated using this setup, but the results presented here are all from the camera with serial number 13890.

\narrowfig{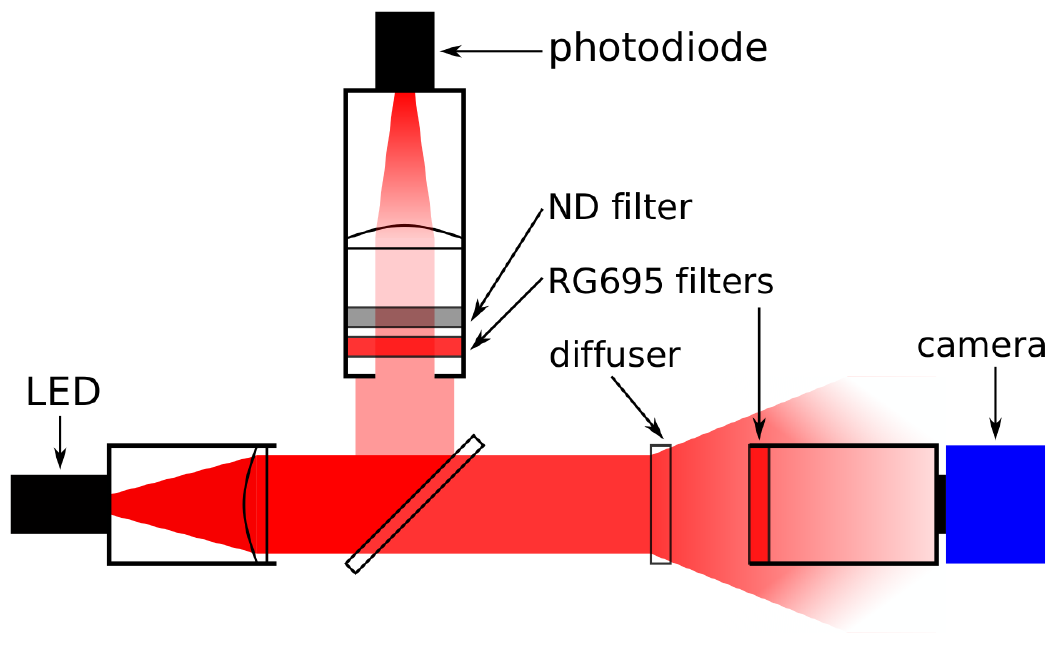}{fig:setup}{Schematic overview of the setup used for the characterization of the A1024B APS sensor.
RG695 colored glass filters and lens tubes are used to reject ambient light.
An uncoated piece of glass is used as a beamsplitter to send about $10\%$ of the light to the photodiode, while the remaining light is sent through a weak ground-glass diffuser to the camera.}

\section{ADC Characterization}\label{sec:biterror}

\widefig{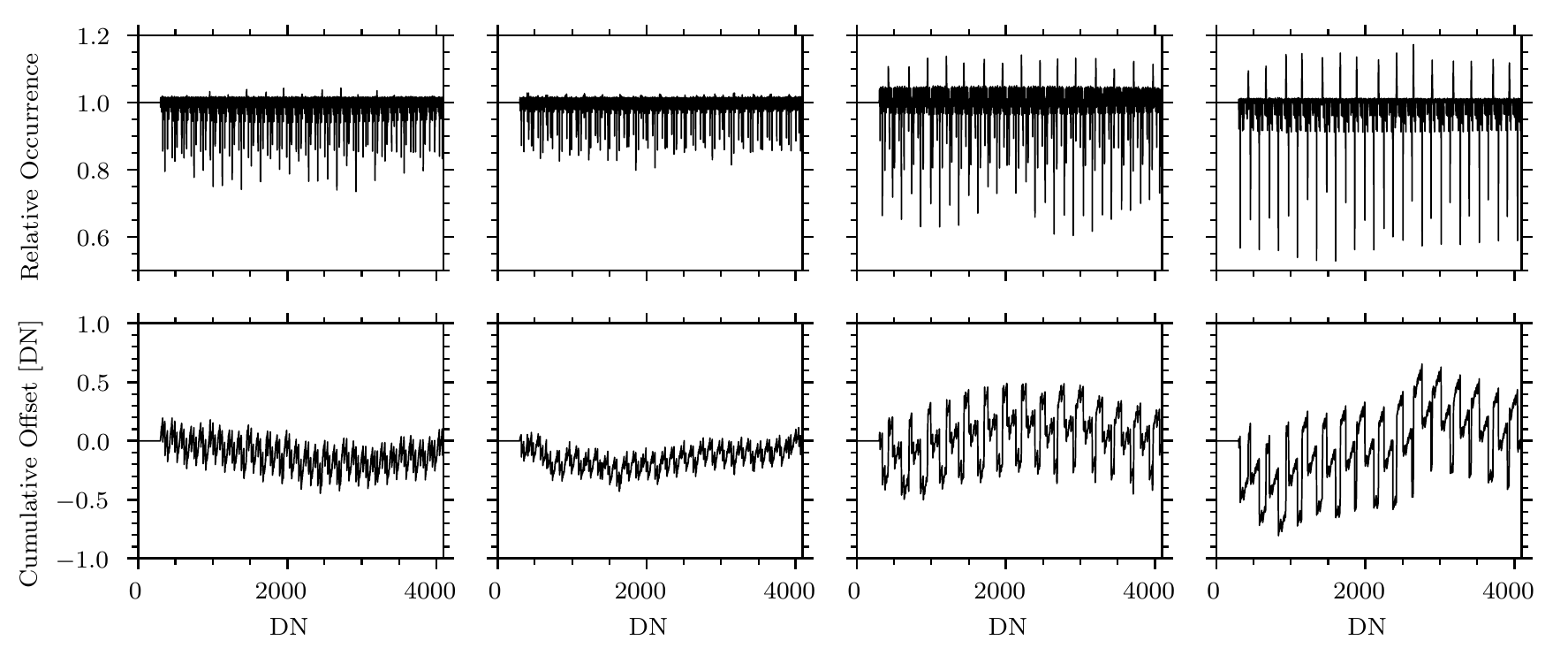}{fig:biterror}{Normalized histograms and cumulative offset of readout values of the four ADCs in a MV-D1024E camera.
The DN error is visible as the difference from unity in the normalized histograms.
The DN error shows clearly periodic behavior as is expected from the architecture of the ADC.}

The MV-D1024E camera uses what is known as a pipelined ADC.
This type of ADC works by successively digitizing the analog signal in blocks of several bits, starting with the most significant one and working toward the least significant one.
In a final step, the results from each step of the ADC are combined to produce an output.
In effect, it combines the speed of flash ADCs and the accuracy of successive-approximation ADCs.
Flash ADCs convert an analog signal in a single step using many comparators.
Successive-approximation ADCs iteratively approach a solution by comparing the analog signal to the output of a digital to analog converter.
Flash ADCs are fast but limited in accuracy because the required number of comparators doubles for every bit in the output, and successive-approximation ADCs are slow but can be made accurate with relative ease.
Pipelined ADCs are frequently used in applications such as video that require high speed and moderate accuracy.
The PhotonFocus MV-D1024E uses four ADCs.
Column $N$ of the sensor is digitized by ADC $N\mod4$.
At a frame rate of $150~\mathrm{Hz}$ each ADC must digitize nearly 40 million samples per second.
ADCs are not designed to produce results that are much more accurate than their least significant bit.
The error is manifested as small differences in the equivalent value of each bit.

The data set for this calibration requires good coverage of all readout values, but no spatial information needs to be retained beyond separation of the ADCs.
The photodiode reference voltage also does not need to be recorded.
The data acquisition procedure recorded histograms of readout values for each of the ADCs in an area near the center of the sensor.
For each camera data were collected at $130, 131, \dots,$ and $139$ levels of illumination, each starting at no illumination and increasing to saturation of the sensor readout.
The histograms of the data are summed and flattened by fitting a spline and dividing by the fit result.
A second flatting step is performed by dividing the histograms by the median over a running window.
The number of points in the spline fit as well as the width of the window over which the median is calculated were adjusted until a satisfactory result was achieved.
The values selected were $50$ and $21$, respectively.
Very low readout values occur infrequently, so they are excluded by setting the first $150$ values of the normalized and flattened histogram to $1$.
The last value, i.e., saturation, is also set to $1$.
Figure~\ref{fig:biterror} shows the resulting histograms for each of the four ADCs in a camera, as well as the cumulative offset of the camera ADCs from a perfect ADC.
The histogram and the cumulative offsets show a clear pattern that is associated with the imperfections of the ADC.

Figure~\ref{fig:biterror} shows that cumulative offsets of about 1~DN can be expected, and the offset may change dramatically over very small ranges of DN values.
For measurements such as polarimetry that rely on accurately determining small differences in illumination, this may lead to large errors.

A model was created to study the effects of DN error on the K-Cor observations.
For simplicity the telescope is assumed to be polarization-free, and the modulator, polarimetric analyzer, and data calibration are all assumed to be perfect.
The telescope is assumed to have an aperture of $17~\mathrm{cm}$, a bandwidth of $30~\mathrm{nm}$ centered at $735~\mathrm{nm}$, and $5\arcsec$ square pixels.
Instrument transmission and detector quantum efficiency (QE) are assumed to be $29\%$ and $30\%$, respectively.
A known input Stokes vector is ``observed'' by applying the modulation matrix, telescope transmission, and detector QE, and subsequently adding a dark offset, yielding a photoelectron count for each state of the modulator.
For each of these a probability mass function (PMF) is constructed consisting of a Poissonian distribution for photon noise convolved with a Gaussian distribution for read noise.
The resulting PMF is multiplied with the cumulative sum of the measured ADC histogram and integrated to yield the mean readout value in DN.
The standard deviation of the readout is computed in a similar fashion.

\narrowfig{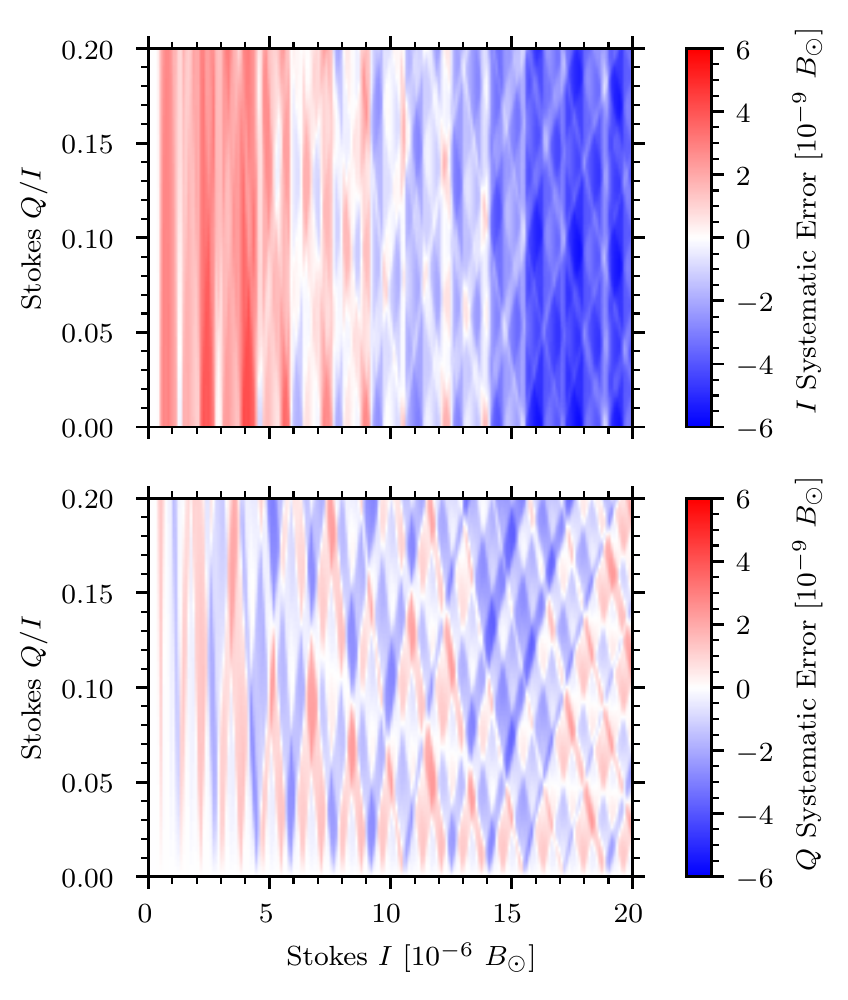}{fig:kcormodel}{Results of a K-Coronagraph model run.
The x and y axes indicate the input Stokes vector, and color indicates the systematic error determined from the simulation.
The input stokes vectors had no U or V component.}

The synthesized observations are reduced by applying dark and gain correction derived from simulated calibration data.
A Stokes vector is recovered through a perfect demodulation derived directly from the known modulation matrix \citep{2000ApOpt..39.1637D}.
Knowledge of systematic errors can now be derived by comparing the observed Stokes vectors against known input Stokes vectors.

Figure~\ref{fig:kcormodel} shows the systematic error found in a simulation using the 4th ADC of the K-Coronagraph camera.
Systematic errors reach several times $10^{-9}~B_\odot$, which is around the desired accuracy of the instrument, about an order of magnitude above the desired sensitivity.
In order to meet the requirements of the instrument the performance of the ADC must be improved, a correction must be applied, or some other way must be found to mitigate the ADC errors.

\narrowfig{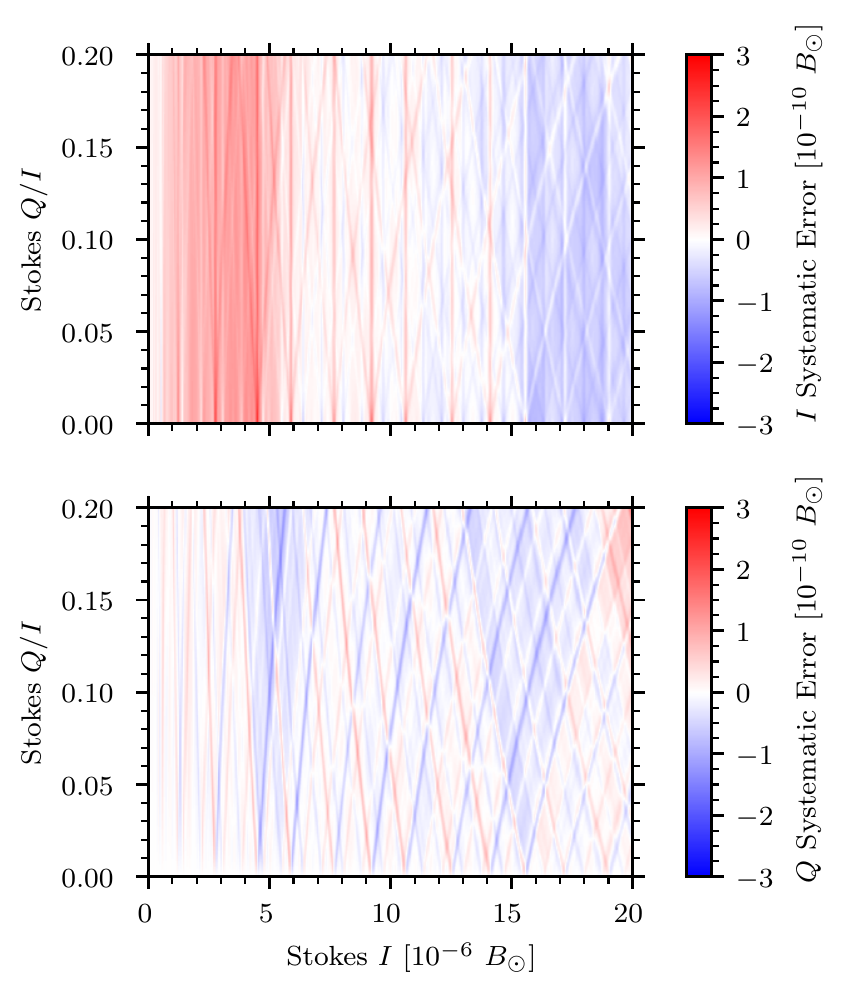}{fig:kcormodellookup}{Like Fig.~\ref{fig:kcormodel} but now using a lookup table with noise to correct DN error.
The systematic error has reduced in amplitude by an order of magnitude.}

The results presented above are for a camera that was modified by the vendor with a 14-bit ADC that is a drop-in replacement of the standard 12-bit ADC.
The least significant 2 bits are not read out, but the camera still benefits from the improved accuracy of the ADC.
There is no other easy way to reduce the ADC errors for the K-Coronagraph cameras.
The GONG project \citep{1996Sci...272.1284H} implemented ramping of the ADC offset to dither the measurement to mitigate similar issues \citep{gongtech7}.
However, such an approach is difficult to implement in modern cameras that have integrated digitization circuitry.
A correction must be applied after the data are digitized.

The most straight-forward solution is to apply a correction in the form of a lookup table, which conveniently consists of the cumulative sum of the relative occurrence of readout values, i.e., the ADC histograms in the top row of Fig.~\ref{fig:biterror}.
For example, if the camera reads out a value of 1000~DN, the lookup table will return the sum of the relative occurence of 0 through 1000~DN.
For the 4th ADC shown in Fig.~\ref{fig:biterror} this yields a value of $1000.6757$.
If the camera reads out a value of 1001~DN, the lookup table will return a value of $1000.6757 + 1.0051 = 1001.6858$, since the relative occurence of 1001~DN is $1.0051$.
In the example here the lookup table uses floating-point numbers.
For the K-Coronagraph an implementation was chosen that uses only integer math.
The lookup table returns a 23-bit integer value that has the effect of adding 11~bits of resolution while allowing the summation of $2^9=512$ frames into a single 32-bit integer without risk of overflow.

The model was modified to include a step where the lookup table is applied.
Since the model uses the ADC histogram to create the error, using the same ADC histogram for the lookup table would result in a perfect correction.
In a real-world scenario, there will be some error on the determination of the lookup table.
To account for this errors here the lookup table is constructed from the measured ADC histogram with $1\%$ added Gaussian noise.
Figure~\ref{fig:kcormodellookup} shows that the systematic error is reduced by about an order of magnitude by this procedure, so that it is now well below the required measurement accuracy.

\section{Pixel-to-Pixel Gain Variation and Nonlinearity}\label{sec:nonlinear}

\narrowfig{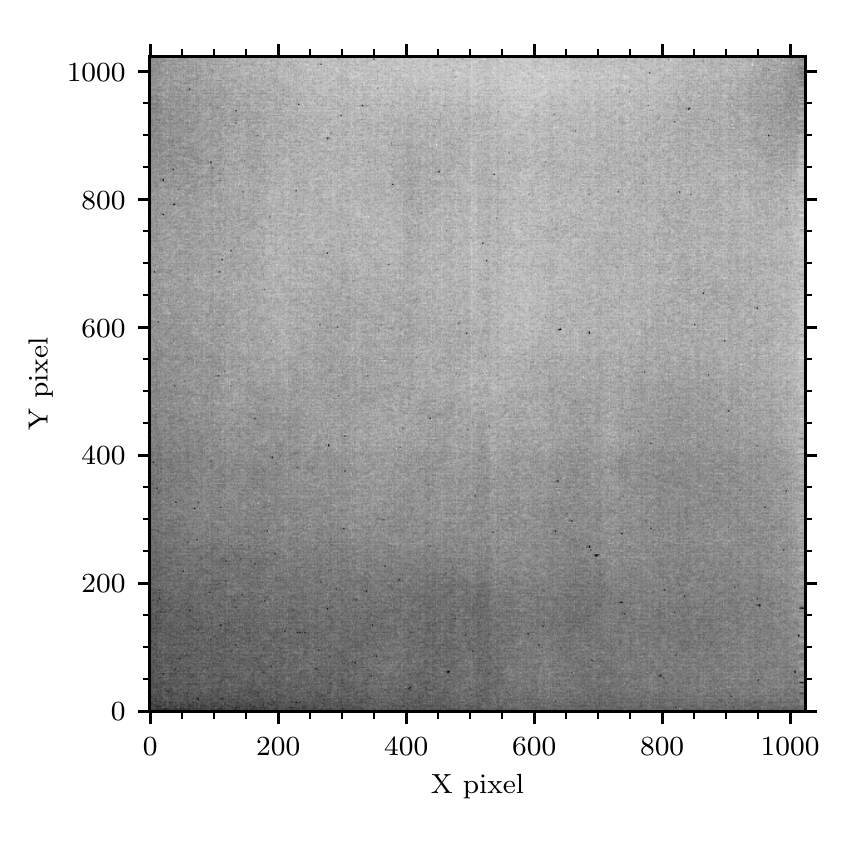}{fig:A1024B_example}{Example raw image taken with a PhotonFocus MV-D1024E camera using the A1024B APS sensor.
The image was taken near $57\%$ saturation of the sensor and is scaled between $47\%$ and $67\%$ saturation.
It is the image that is used as the ``flat'' in the dark and gain correction described in the text.}

\widefig{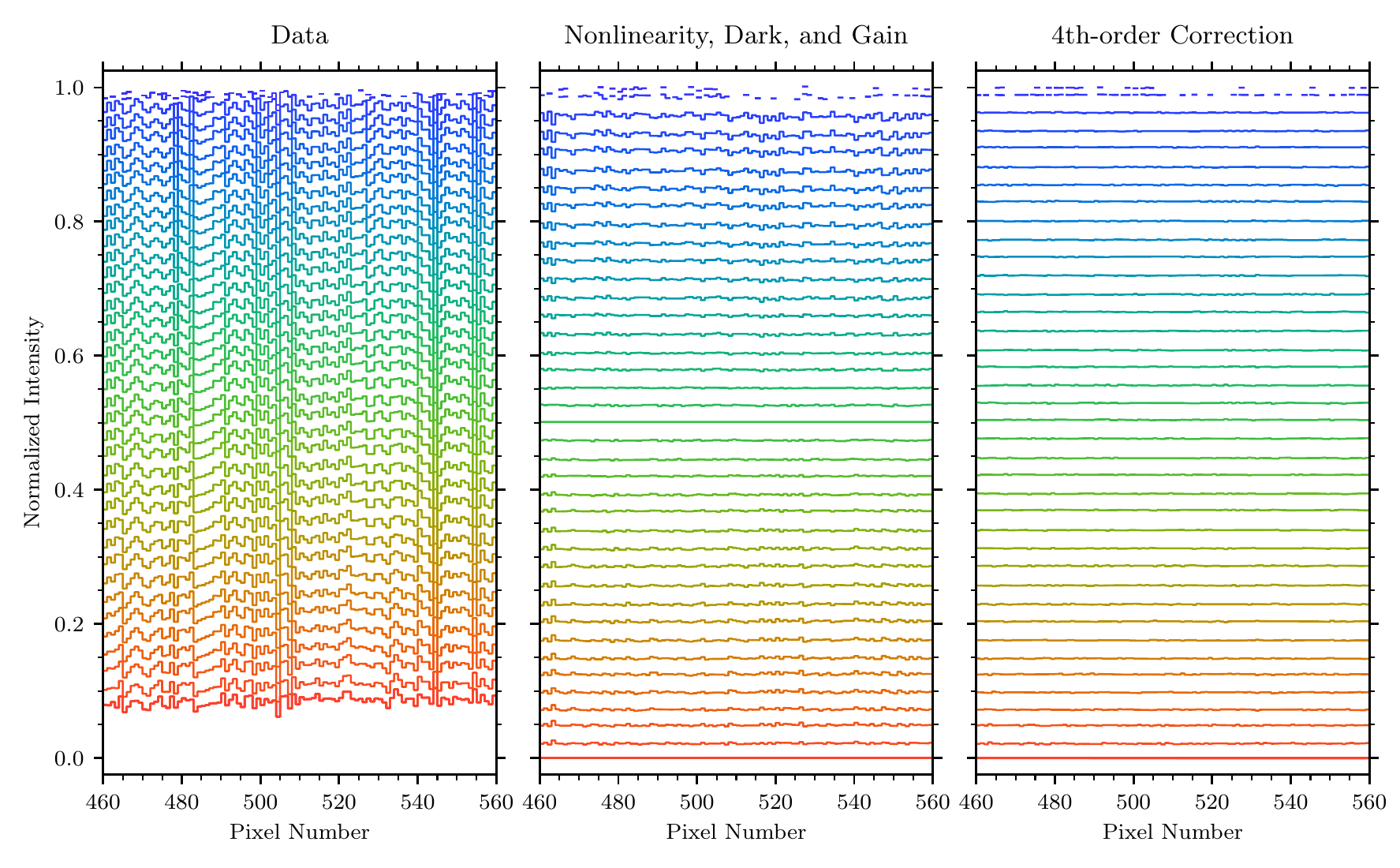}{fig:biasgain}{Comparison of the raw data (left), standard correction with a global nonlinearity, bias, and gain (middle), and a 4th-order polynomial correction to each pixel.
The plots show the normalized averaged intensity value for 100 pixels near the middle of the detector at 40 illumination levels.
The raw data show structure with amplitudes around $0.02$ that is highly correlated between illumination levels.
The standard correction uses the unilluminated exposure as the dark frame and 22nd level as the gain.
These levels consequently show a perfect correction.
Levels above and below show highly correlated structure that increases in amplitude as the illumination level increases in difference to the dark and gain levels.
The 4th-order fit shows substantially improved correction at those levels.
The interruption of the lines at high illumination level is due the omission of pixels that are saturated.}

The A1024B APS sensor exhibits substantial pixel-to-pixel variation.
An example image at approximately $57\%$ saturation is shown in Fig.~\ref{fig:A1024B_example}.
Because this variation is systematic it cannot be reduced through averaging of frames, but instead must be removed through the application of a correction.
Traditionally, the correction applied consists of the subtraction of a dark frame and subsequent division by a gain image.
The gain image can be constructed by subtracting the dark image from an image taken with flat illumination.
In the case of the K-Coronagraph flat illumination is achieved by inserting an opal diffuser in front of the objective.
Other methods of deriving a gain image exist such as the iterative method devised by \cite{1991PASP..103.1097K} and derived methods \citep[e.g.,][]{2004SoPh..221....1C, 2016ApJ...827..137X}.

In addition to gain variation both CMOS and CCD sensors and camera readout electronics usually exhibit nonlinear response to incident light that must be corrected for the data to be of scientifically useful quality, particularly if those data are intended for quantitative analysis such as polarimetry.
The double-aperture method has been used to accurately measure nonlinear response \citep{elster1913,1972ApOpt..11.2294M,2012Metro..49S..99T}.
However, such sophisticated characterization setups are not typically available, and it is desirable to have other, simpler methods.

A common approach is to have constant illumination with varying exposure time.
While this method is usually comparatively straight-forward to implement \citep[e.g.,][]{1993AJ....106.2441G}, it requires that the exposure timing is accurate and that the camera is well-behaved over several orders of magnitude of exposure time.
Standard 1288 of the European Machine Vision Association\footnote{\url{https://www.emva.org/standards-technology/emva-1288/}} suggests continuous but variable illumination and pulsed illumination with constant exposure time as methods to control sensor irradiation.
Others have used a fixed linear polarizer together with a rotating one to create predictable modulation of incident light \citep[e.g.,][]{2011ApOpt..50.2401H}.
Here, the camera was calibrated against a photodiode that is known to have very linear response (see Sect.~\ref{sec:setup}).

For this calibration full frame images were recorded with the calibration setup at 41 levels of illumination, starting at no illumination and increasing to saturation of the sensor readout.
The photodiode reference voltage was recorded simultaneously.
At each level of illumination $256$ frames were collected.
For each pixel, the average value as well as the standard deviation were calculated.
For the camera tested here 63 pixels with low intensities or anomalous trends with respect to the photodiode voltage are masked out and excluded from the analysis.
Since no calibration will be available for these pixels they will need to be replaced in the science data products, e.g., with the average of the surrounding pixels.
Additionally, measurements that are affected by saturation are also excluded.
The leftmost panel of Fig.~\ref{fig:biasgain} shows example raw data with large systematic variations between pixels.

The traditional dark and gain correction was applied by using the data at no illumination as the dark signal and the frame at approximately half saturation of the sensor as the flat signal.
A global nonlinearity was fit with a 3rd-order polynomial and corrected prior to the application of the dark and gain correction.
The result of this procedure is shown in the middle panel of Fig.~\ref{fig:biasgain}.
There is clear improvement over the raw data, and the correction is very good near the level of the flat signal.
However, significant systematic errors persist, as can be seen from highly correlated errors at different levels of illumination, particularly above a normalized intensity of $0.8$.
The RMS noise is shown by the solid orange curve in Fig.~\ref{fig:noise}.
It is below $10^{-2}$ at all illumination levels, and shows a correction to $10^{-3}$ around the level of the flat signal.

\narrowfig{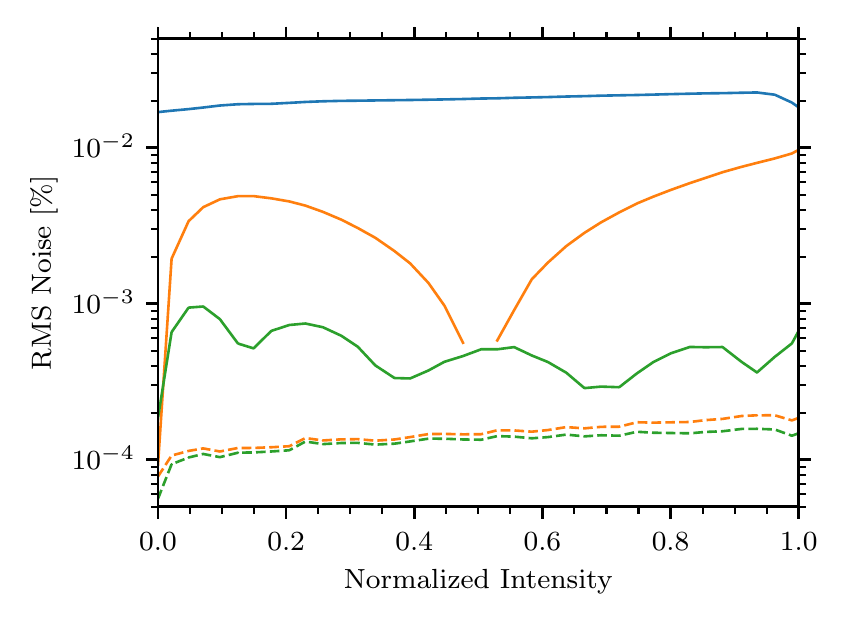}{fig:noise}{Solid curves show the RMS noise as a function of normalized intensity for the raw data (blue), the standard correction (orange), and the 4th-order fit (green).
Dashed curves show the random noise from propagation of uncertainty using the same color scheme.
The RMS noise in the raw data decreases at high illumination levels due to saturation effects.
The standard correction RMS noise values are omitted at the levels used for the dark and the flat data because they are identically zero.
The 4th order fit outperforms the standard correction by up to an order of magnitude everywhere except at illumination levels very near the dark level and the gain level.
Both corrections remain well above the theoretical performance based on random noise propagation, indicating the noise is dominated by an imperfect correction rather than random sources.}

The random noise can be estimated by propagating the standard deviation calculated from the raw data through the analysis.
It ranges from approx.\ $1\times10^{-4}$ to $2\times10^{-4}$ (see Fig.~\ref{fig:noise}).
This further reinforces the conclusion that the RMS noise measurement is dominated by uncorrected systematic errors deduced from the behavior observed in Fig.~\ref{fig:biasgain}.

The data were analyzed again, and a 4th-order polynomial function of the normalized photodiode voltage was fit as a nonlinearity correction for each pixel individually in an effort to improve the correction.
The result of applying this correction is shown in the rightmost panel of Fig.~\ref{fig:biasgain}.
It shows clear improvement over the traditional correction, which is confirmed by calculation of the RMS noise, shown by the solid green curve in Fig.~\ref{fig:noise}.
The RMS noise is now below $10^{-3}$ at all illumination levels, though like the traditional correction, it is still well above the estimated random noise level and thus dominated by uncorrected systematic errors.
Increasing the order of the polynomial does not yield better results.

The pixel-wise correction is a replacement of the global nonlinearity correction and must be used in conjunction with the traditional dark and gain correction.
Dark current depends on sensor temperature and is not expected to be stable with these cameras in the K-Coronagraph.
The gain correction also removes artifacts introduced by the telescope and instrument optics, such as vignetting, dust on surfaces near focal planes, etc.
While the traditional correction yields results that would appear to meet the science requirements of the K-Coronagraph, the systematic errors left are prominent due to very low random noise.
The pixel-wise correction produces noticeably better data quality at a relatively small computational cost.

\section{Conclusion}\label{sec:conclusion}
The selection of the camera for the COSMO K-Coronagraph was driven by the need for a high product of full-well capacity and frame rate.
The PhotonFocus MV-D1024E camera that was selected is intended for industrial vision and it was found that it is not directly suitable for application in the K-Coronagraph without characterization and mitigation of errors introduced by camera hardware.

I discussed the effect of discretization on the averaging of data in Sect.~\ref{sec:discretization}.
The K-Coronagraph camera has adequate bit depth and noise to ensure that data can be averaged well below the required sensitivity of the instrument.
Cameras that reach a similar rate of photoelectrons per second as the MV-D1024E with shallower wells are forced to trade readout bit depth to reach higher frame rates.
Particularly for cameras with low read noise, quantization error can be substantial.

During the initial testing of the MV-D1024E camera it was discovered that the ADC suffers from inaccuracies that manifest themselves as small differences in the equivalent value of each bit.
Through modeling, I showed that this results in systematic errors that are around the level of accuracy required for the K-Coronagraph.
They can be effectively mitigated through the post-facto application of a lookup table.

Lastly, I analyzed the performance of the traditional global nonlinearity correction followed by dark subtraction and division by a gain image.
This procedure yields residual systematic errors of up to $1\%$ of the signal level.
Uniformity is a well-known weakness of CMOS APS sensors.
A pixel-wise nonlinearity correction lowers the residuals below $0.1\%$.

The correction of DN errors is implemented in the K-Coronagraph instrument data acquisition system through a lookup table that is applied to each frame that is recorded prior to summation into a final data product.
The pixel-wise nonlinearity correction was recently implemented in the data reduction pipeline when the first set of cameras was replaced.
Since originally only a global nonlinearity correction was implemented, the old cameras must be re-characterized so that the new correction can be applied to the data that were acquired with them.
An effort is currently underway to perform the characterization and to re-process old data.

While in this paper the focus is on the PhotonFocus MV-D1024E camera the systematic errors described here can be found in many APS CMOS and CCD cameras, and they illustrate the care that must be taken in selecting a suitable camera for an application, and that these errors can be effectively mitigated through relatively simple procedures.

\acknowledgments{I would like to thank S.~Sewell for assistance collecting the calibration data.}

\let\clearpage\tmpclearpage


\begin{thebibliography}{19}
\expandafter\ifx\csname natexlab\endcsname\relax\def\natexlab#1{#1}\fi

\bibitem[{{Boyle} \& {Smith}(1970)}]{1970BSTJ...49..587B}
{Boyle}, W.~S. \& {Smith}, G.~E. 1970, Bell Syst. Tech. J, 49, 587

\bibitem[{{Budde}(1979)}]{1979ApOpt..18.1555B}
{Budde}, W. 1979, \ao, 18, 1555

\bibitem[{{Casini} {et~al.}(2012){Casini}, {de Wijn}, \&
  {Judge}}]{2012ApJ...757...45C}
{Casini}, R., {de Wijn}, A.~G., \& {Judge}, P.~G. 2012, \apj, 757, 45

\bibitem[{{Chae}(2004)}]{2004SoPh..221....1C}
{Chae}, J. 2004, \solphys, 221, 1

\bibitem[{{Chamberlain}(1969)}]{1969IJSSC...4..333C}
{Chamberlain}, S.~G. 1969, IEEE Journal of Solid-State Circuits, 4, 333

\bibitem[{{del Toro Iniesta} \& {Collados}(2000)}]{2000ApOpt..39.1637D}
{del Toro Iniesta}, J.~C. \& {Collados}, M. 2000, \ao, 39, 1637

\bibitem[{Elster \& Geitel(1913)}]{elster1913}
Elster, J. \& Geitel, H. 1913, Phys. Zeit., 14, 741

\bibitem[{{Gilliland} {et~al.}(1993){Gilliland}, {Brown}, {Kjeldsen},
  {McCarthy}, {Peri}, {Belmonte}, {Vidal}, {Cram}, {Palmer}, {Frandsen},
  {Parthasarathy}, {Petro}, {Schneider}, {Stetson}, \&
  {Weiss}}]{1993AJ....106.2441G}
{Gilliland}, R.~L., {Brown}, T.~M., {Kjeldsen}, H., {et~al.} 1993, \aj, 106,
  2441

\bibitem[{{Hanaoka} {et~al.}(2011){Hanaoka}, {Suzuki}, \&
  {Sakurai}}]{2011ApOpt..50.2401H}
{Hanaoka}, Y., {Suzuki}, I., \& {Sakurai}, T. 2011, \ao, 50, 2401

\bibitem[{{Harvey} \& the GONG Instrument Development~Team(1990)}]{gongtech7}
{Harvey}, J. \& the GONG Instrument Development~Team. 1990, Prototype Design
  Review, Tech. Rep.~7, Global Oscillation Network Group,
  \url{http://gong.nso.edu/science/papers/Report\_7.pdf}

\bibitem[{{Harvey} {et~al.}(1996){Harvey}, {Hill}, {Hubbard}, {Kennedy},
  {Leibacher}, {Pintar}, {Gilman}, {Noyes}, {Title}, {Toomre}, {Ulrich},
  {Bhatnagar}, {Kennewell}, {Marquette}, {Patron}, {Saa}, \&
  {Yasukawa}}]{1996Sci...272.1284H}
{Harvey}, J.~W., {Hill}, F., {Hubbard}, R.~P., {et~al.} 1996, Science, 272,
  1284

\bibitem[{{Hou} {et~al.}(2013){Hou}, {de Wijn}, \&
  {Tomczyk}}]{2013ApJ...774...85H}
{Hou}, J., {de Wijn}, A.~G., \& {Tomczyk}, S. 2013, \apj, 774, 85

\bibitem[{{Kuhn} {et~al.}(1991){Kuhn}, {Lin}, \&
  {Loranz}}]{1991PASP..103.1097K}
{Kuhn}, J.~R., {Lin}, H., \& {Loranz}, D. 1991, Publications of the
  Astronomical Society of the Pacific, 103, 1097

\bibitem[{{Mielenz} \& {Eckerle}(1972)}]{1972ApOpt..11.2294M}
{Mielenz}, K.~D. \& {Eckerle}, K.~L. 1972, \ao, 11, 2294

\bibitem[{{Noble}(1968)}]{1968ITED...15..202N}
{Noble}, P.~J.~W. 1968, IEEE Transactions on Electron Devices, 15, 202

\bibitem[{{Sewell} \& {Tomczyk}(2010)}]{cosmotech17}
{Sewell}, S. \& {Tomczyk}, S. 2010, Photonfocus Camera Evaluation, Tech.
  Rep.~17, Coronal Solar Magnetism Observatory

\bibitem[{{Theocharous}(2012)}]{2012Metro..49S..99T}
{Theocharous}, E. 2012, Metrologia, 49, S99

\bibitem[{{Weimer} {et~al.}(1969){Weimer}, {Pike}, {Sadasiv}, \&
  {Shallcross}}]{1969ITED...16..240W}
{Weimer}, P.~K., {Pike}, W.~S., {Sadasiv}, G., \& {Shallcross}, F.~V. 1969,
  IEEE Transactions on Electron Devices, 16, 240

\bibitem[{{Xu} {et~al.}(2016){Xu}, {Zheng}, {Lin}, \&
  {Wang}}]{2016ApJ...827..137X}
{Xu}, G.-G., {Zheng}, S., {Lin}, G.-H., \& {Wang}, X.-F. 2016, \apj, 827, 137

\end{thebibliography}
\end{document}